\documentclass[aps,prd,amssymb,final,nobibnotes,nofootinbib,twocolumn,superscriptaddress]{revtex4}
\usepackage{graphicx}
\usepackage{hyperref}
\usepackage{amsmath}
\usepackage{color}
\usepackage{xcolor}
\usepackage{numprint}
\usepackage{float}
\usepackage{ulem}
\usepackage{physics}
\usepackage{mathtools}
\usepackage{dsfont}
\usepackage{subfigure}
\usepackage{upgreek}

\usepackage{lipsum}

\begin{document}
\title{Radiation emitted by a source orbiting a Schwarzschild--anti--de Sitter black hole }

\author{Jo\~ao P. B. Brito}
\email{joao.brito@icen.ufpa.br} 
\affiliation{Programa de P\'os-Gradua\c{c}\~{a}o em F\'{\i}sica, Universidade 
		Federal do Par\'a, 66075-110, Bel\'em, Par\'a, Brazil.}

\author{Rafael P. Bernar}
\email{rbernar@ufpa.br}
\affiliation{Programa de P\'os-Gradua\c{c}\~{a}o em F\'{\i}sica, Universidade 
		Federal do Par\'a, 66075-110, Bel\'em, Par\'a, Brazil.}

\author{Lu\'{\i}s C. B. Crispino}
\email{crispino@ufpa.br} 
\affiliation{Programa de P\'os-Gradua\c{c}\~{a}o em F\'{\i}sica, Universidade 
		Federal do Par\'a, 66075-110, Bel\'em, Par\'a, Brazil.}

\date{\today}
\begin{abstract}
  We study the scalar radiation emitted by a source orbiting a Schwarzschild--anti--de Sitter black hole in circular geodesics. We obtain the emitted power using quantum field theory in curved spacetimes at tree level with reflective boundary conditions at infinity. We compare our results with the ones obtained in Schwarzschild and Schwarzschild--de Sitter ($\Lambda > 0$) spacetimes.
  In the Schwarzschild-de Sitter spacetime we observe an enhancement of the emitted power associated to the lower multipole modes, when compared with the Schwarzschild case. In contrast, for the Schwarzschild-anti-de Sitter spacetime, the enhancement is rather of the higher multipole modes.

\end{abstract}


\maketitle

\section{Introduction}

The detections of gravitational waves from black hole (BH) binary mergers by the Ligo and Virgo Scientific Collaborations~\cite{ligo1_2016,ligo2_2016}, together with the first image of a BH shadow~\cite{EHT_sombra}, mark the beginning of a new era for gravitational physics. In particular, the physics around BHs has become even more important, since these phenomena may now be better constrained by the experimental data available.
BH spacetimes possess a nontrivial causal structure and their strong gravitational field implies that they are important pieces not only in General Relativity (GR), but also in alternative theories of gravity~\cite{wald_1984,berti_2015}.
On one hand, the classical aspects of BH interactions with matter have importance in the astrophysical context, since these compact objects are believed to exist in the center of almost all galaxies, being one of the main sources of power in them~\cite{Hoyle1963,Salpeter1964,Lynden-Bell1969,Kormendy2013}. 
On the other hand, BHs causal structure provides an interesting setting to study the quantum aspects of matter, particularly fields, since we can use the quantum field theory (QFT) in curved spaces framework. The theoretical implications of this analysis for the quantum fields are interesting in their own right, but they might also provide insights to the quantum nature of gravity.  

Ever since its formulation more than a hundred years ago, GR has passed several experimental tests (see Ref.~\cite{Turyshev2008} for a review) and is regarded as the established theory to describe the physics of spacetime. However, it is a classical theory plagued with singularities~\cite{hawking_1996}, which breaks down in the Planck scale, such as in the very beginning of our universe or inside a BH. A satisfactory quantum theory of gravity is still missing, being one outstanding issue in fundamental physics~\cite{kiefer_2005}. In an effort to understand possible quantum aspects of gravity, we may use an effective approach, in which quantum fields are coupled to the gravitational field represented by a fixed classical background spacetime~\cite{birrel_1982,parker_2009}. Even if QFT in curved spacetimes is not a fundamental theory, important results have been obtained using this framework, namely the particle creation in expanding universes~\cite{parker_1969} and BH spacetimes (Hawking radiation)~\cite{hawking_1975}, as well as the Unruh effect, which establishes the observer-dependent nature of the particle content in field theory~\cite{unruh_1976}. Hence, although being an effective theory, QFT in curved spacetimes sheds light on important aspects of the interface between gravity and quantum physics.

Accretion disks are expected to surround most BHs found in the Universe. As matter spirals around these compact objects, it emits radiation through several channels. As previously mentioned, BHs and their surrounding matter are generally a main source of power in galaxies and the associated radiation processes play an important role in high-energy astrophysics.
QFT may be used to compute radiation emission by accelerated sources in flat~\cite{alves_2004,alves_2010,castineiras_2011} and curved spacetimes~\cite{castineiras_2002}. 
In the context of radiation processes near BHs, QFT can be used to analyze the interaction of sources with Hawking radiation~\cite{crispino_1998} and the radiation emission by sources in geodesic motion.
We may analyze these phenomena by considering a source moving along circular geodesics. For unstable orbits, there is the possibility of radiation emission of the synchrotron type, hence this emission is usually called the geodesic synchrotron radiation. This was initially analyzed for a scalar field in the Schwarzschild background using the Green function framework~\cite{misner_1972,misner_et_al_1972}. We can treat this phenomenon with a semiclassical approach using the QFT in curved spacetimes framework. The scalar radiation was investigated in asymptotically flat spacetimes, e. g., in Refs.~\cite{crispino_2000,castineiras_2007,crispino_2008,crispino_2009,macedo_2012,bernar_2019}, the electromagnetic radiation was studied in Ref.~\cite{castineiras_2005} and the gravitational radiation was analyzed in Refs.~\cite{bernar_2017,bernar_2018}. In the context of spacetimes with nonvanishing cosmological constant, the scalar radiation was studied in the Schwarzschild--de Sitter (SdS) spacetime in Ref.~\cite{brito_2020} and in the Schwarzschild--anti--de Sitter (SAdS) spacetime, using the Green function method, in Ref.~\cite{cardoso_2002}. In the SAdS spacetime an investigation of the geodesic scalar radiation using QFT in curved spacetimes has not been previously carried out. 
In particular, one motivation to study dynamical processes in asymptotically anti--de--Sitter spacetimes is concerned with the conjectured anti--de--Sitter/conformal field theory (AdS/CFT) correspondence, in which physics in the bulk of AdS is related to the physics in the conformal field theory boundary (see, e.g., Ref.~\cite{aharony_2000} and the references therein).

Solutions of the GR field equations with a nonvanishing cosmological constant are not asymptotically flat. In particular, for the SAdS solution ($\Lambda < 0$), in contrast to an asymptotically flat solution, the spacetime boundary at spatial infinity is a timelike surface. 
The SAdS solution describes a nonglobally hyperbolic spacetime and, therefore, does not possess a Cauchy surface on which one can give initial data. In order to have a well defined QFT, we must impose suitable boundary conditions~\cite{isham_1978,morley_2020,ishibashi_2004,wald_1980}. In this paper, we use QFT in curved spacetimes at tree level to analyze the scalar radiation emitted by a source in geodesic circular motion around a SAdS BH, considering reflective boundary conditions at infinity.

The remainder of this paper is organized as follows. In Sec.~\ref{sec_SdS_black_hole}, we review some general features of the SAdS spacetime, analyzing the geodesic structure and its associated conserved quantities. In Sec.~\ref{sec_scalar_radiation}, we study the scalar field dynamics in the curved background and we canonically quantize the field, choosing reflective boundary conditions at $r \to \infty,$ where the field effective potential diverges. In Sec.~\ref{Sec_radiation}, we compute the power emitted  by the rotating source considering first order in perturbation theory and using numerical methods to solve the differential wave equation. In Sec.~\ref{sec_results}, we exhibit some selected results and we present our final remarks in Sec.~\ref{Sec_remarks}. In this paper, we adopt natural units such that $c=G=\hbar=1$ and the metric signature ($-,+,+,+$).

\section{Schwarzschild--anti--de Sitter black holes}
\label{sec_SdS_black_hole}
In this section, we present some important properties of the SAdS spacetime, which is a spherically symmetric vacuum solution of Einstein field equations with negative cosmological constant, characterized by a central black hole with geometric mass $M.$ The SAdS line element is given by~\cite{stuchlik_1999}
\begin{equation}
\label{SAdS_line_element}
ds^2 = -f_{\Lambda}(r)dt^2 + \frac{dr^2}{f_{\Lambda}(r)} + r^2(d\theta^2 + \sin^2 \theta d\phi^2),
\end{equation}
with
\begin{equation}
\label{f}
f_{\Lambda}(r) \equiv 1 - \frac{2 M}{r} - \frac{\Lambda}{3}r^2.
\end{equation}
The line element, given by Eq.~\eqref{SAdS_line_element}, is static for $r>r_e,$ where $r_e$ is the radial location of the black hole event horizon ($H_e$). The radial position $r_e$ is obtained by solving
\begin{equation}
\label{f_zero}
f_{\Lambda}(r_{e}) = 0,
\end{equation}
from which the real solution is given by
\begin{equation}
\label{event_horizon}
r_e = -\frac{\Lambda + \zeta^{2/3}}{\Lambda \zeta^{1/3}},
\end{equation}
where
\begin{equation}
\label{defined_function}
\zeta \equiv \sqrt{\Lambda^3 \left(9 M^2 \Lambda -1 \right)} + 3 M \Lambda^2.
\end{equation}
Another useful quantity is the so-called anti-de Sitter radius, which is the scale associated with the cosmological constant and given by
\begin{equation}
R_{AdS} = \sqrt{\frac{3}{\abs{\Lambda}}}.
\end{equation}

\begin{figure}
\includegraphics[scale=0.42]{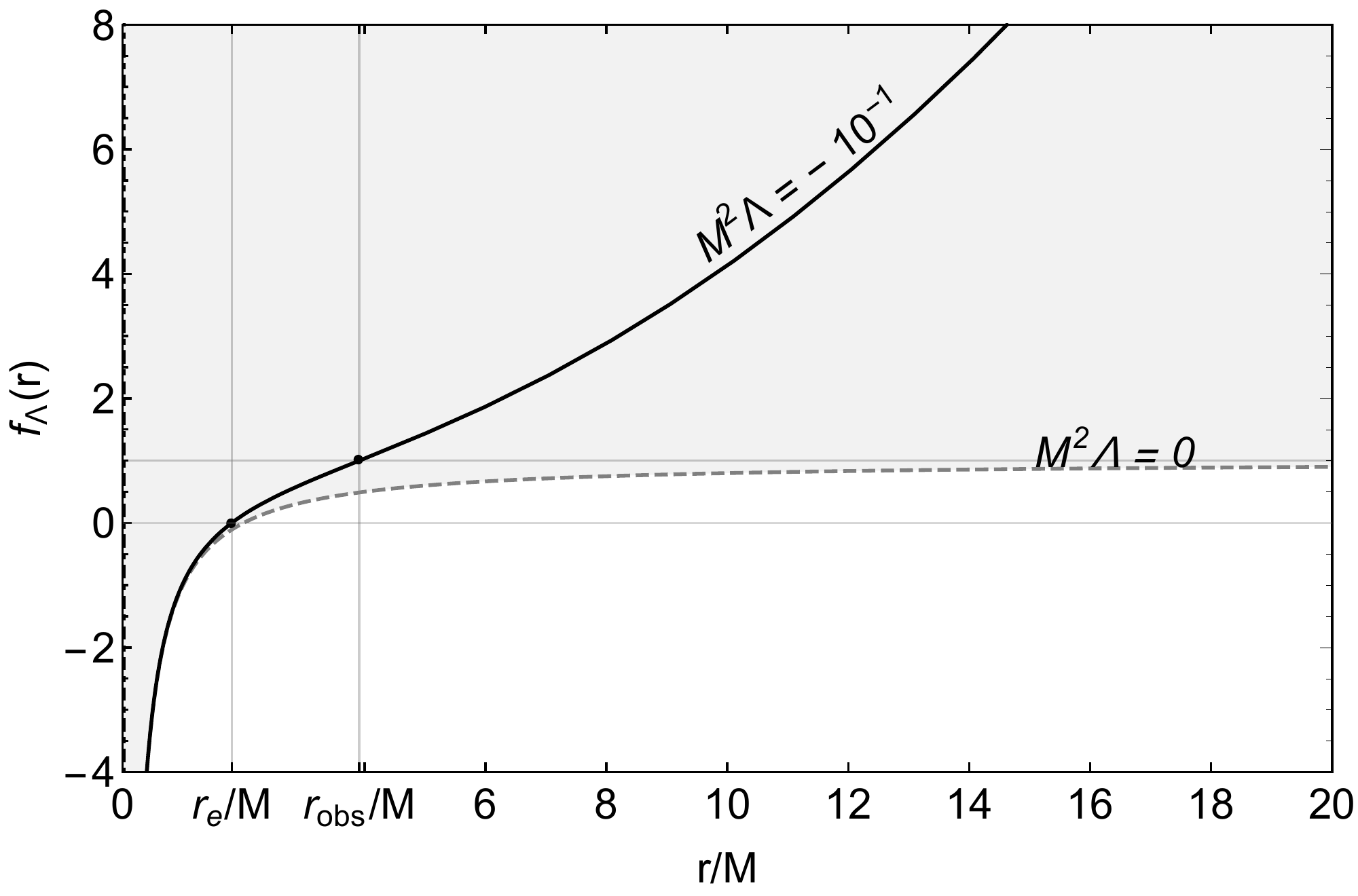}
\caption{The function $f_{\Lambda}(r),$ given by Eq. (\ref{f}). The gray region covers all possible values of $\Lambda$ in the interval $-\infty < \Lambda < 0.$}
\label{fig_radial_function}
\end{figure}

The Schwarzschild solution is obtained in the limit $\Lambda \rightarrow 0,$ in Eq.~\eqref{f}, for which $r_e \rightarrow 2M.$ The anti--de Sitter (AdS) solution~\cite{hawking_1973} is obtained in the limit $M \rightarrow 0,$ for which $r_e \rightarrow 0.$ 
For a fixed nonvanishing $M$, in the limit $\Lambda \rightarrow - \infty$ we have that  $r_e \rightarrow 0$.
The behavior of $f_{\Lambda}(r)$ is illustrated in Fig.~\ref{fig_radial_function}.

We shall consider a source that orbits the black hole along circular geodesics in the equatorial plane. 
The choice of motion in the equatorial plane can be done, without loss of generality, due to the spherical symmetry of the SAdS spacetime.
In the next subsection, we review some features of the SAdS circular geodesics.

\subsection{Geodesics in SAdS spacetime}
The dynamics of test-particles in a curved background can be derived from the Lagrangian
\begin{equation}
\label{lagrangian}
\mathcal{L}_P = \frac{1}{2}g_{\mu \nu} \dot{x}^{\mu} \dot{x}^{\nu},
\end{equation}
where the overdot denotes differentiation with respect to an affine parameter (in the case of timelike paths, we identify the affine parameter as the particle's proper time). The components of the metric $g_{\mu \nu}$ are obtained from Eq.~(\ref{SAdS_line_element}).

There are two integrals of motion along the geodesics in the equatorial plane ($\theta = \pi/2$ and $\dot{\theta} = 0$) due to the time and axial Killing vectors, $\partial_t$ and $\partial_{\phi},$ respectively.
The corresponding conserved quantities can be written as
\begin{eqnarray}
\label{E_circular}
P_t &=& -\frac{\partial \mathcal{L}_P}{\partial \dot{t}} = f_{\Lambda}(r) \dot{t} \equiv E, \label{integral_motion_E}\\
\label{L_circular}
P_{\phi} &=& \frac{\partial \mathcal{L}_P}{\partial \dot{\phi}} = r^2 \dot{\phi} \equiv L. \label{integral_motion_L}
\end{eqnarray}
The quantities $E$ and $L$ are identified as the specific energy and angular momentum of the test-particle, as seen by a static observer at the radial position $r_{obs} \equiv (6M/\abs{\Lambda})^{1/3}.$ Note that $r_{obs} \rightarrow \infty$ as $\abs{\Lambda} \rightarrow 0.$ 

Given that $2 \mathcal{L}_P \equiv \epsilon = -1$ ($0$) for timelike (null) geodesics, using Eqs.~\eqref{integral_motion_E} and \eqref{integral_motion_L} one finds that the test-particle motion is determined by the first order non-linear differential equation:
\begin{equation}
\label{energy}
\dot{r}^2 = E^2 - 2V_P(r),
\end{equation}
where we have defined the central potential as
\begin{equation}
\label{central_potenctial}
V_P(r) \equiv \frac{1}{2}f_{\Lambda}(r) \left(-\epsilon + \frac{L^2}{r^2} \right).
\end{equation}
For timelike geodesics, the potential $V_P(r)$ presents points of maximum and minimum. It vanishes at the black hole horizon $r_e$ and diverges at infinity ($r=\infty$). For null geodesics, the potential $V_P(r)$ has a maximum at the radial position $r_0 \equiv 3M$ (unstable photon sphere). It vanishes at the event horizon and tends to $-L^{2}\Lambda/6$ [$= V_P(2M)$] at infinity. Accordingly, massless particles can escape to infinity, in contrast to massive particles, which are confined due to the diverging potential barrier as $r \to \infty.$

Let us analyze circular geodesics that are constrained by the conditions $\dot{r} = 0$ and $\ddot{r} =0.$ From Eqs.~\eqref{energy} and \eqref{central_potenctial}, we obtain the conserved quantities along the timelike circular geodesics:
\begin{equation}
\label{circular_E_L}
E^2 = r\frac{f_{\Lambda}(r)^2}{r-3M}, \hspace{1 cm} L^2 = r^2 \frac{M - r^3 \Lambda /3}{r - 3M}.
\end{equation}
Noting that $E$ and $L$ must be real quantities, it follows that circular geodesics must exist in the region
\begin{equation}
\label{circular_range}
3M < r < +\infty.
\end{equation}

The lower limit of the interval given by Eq.~\eqref{circular_range} is the radial position of the circular null geodesic, $r_0$, which is independent of $\Lambda.$

By analyzing the central potential given by Eq.~\eqref{central_potenctial}, it is straightforward to obtain the condition that stable circular orbit parameters in the SAdS spacetime must obey, namely
\begin{equation}
\label{stable_condition}
S(\Lambda,r) \equiv -4 \Lambda r^4 + 15 \Lambda M r^3 + 3Mr - 18 M^2 \geq 0.
\end{equation}

\begin{figure}
\includegraphics[scale=0.4]{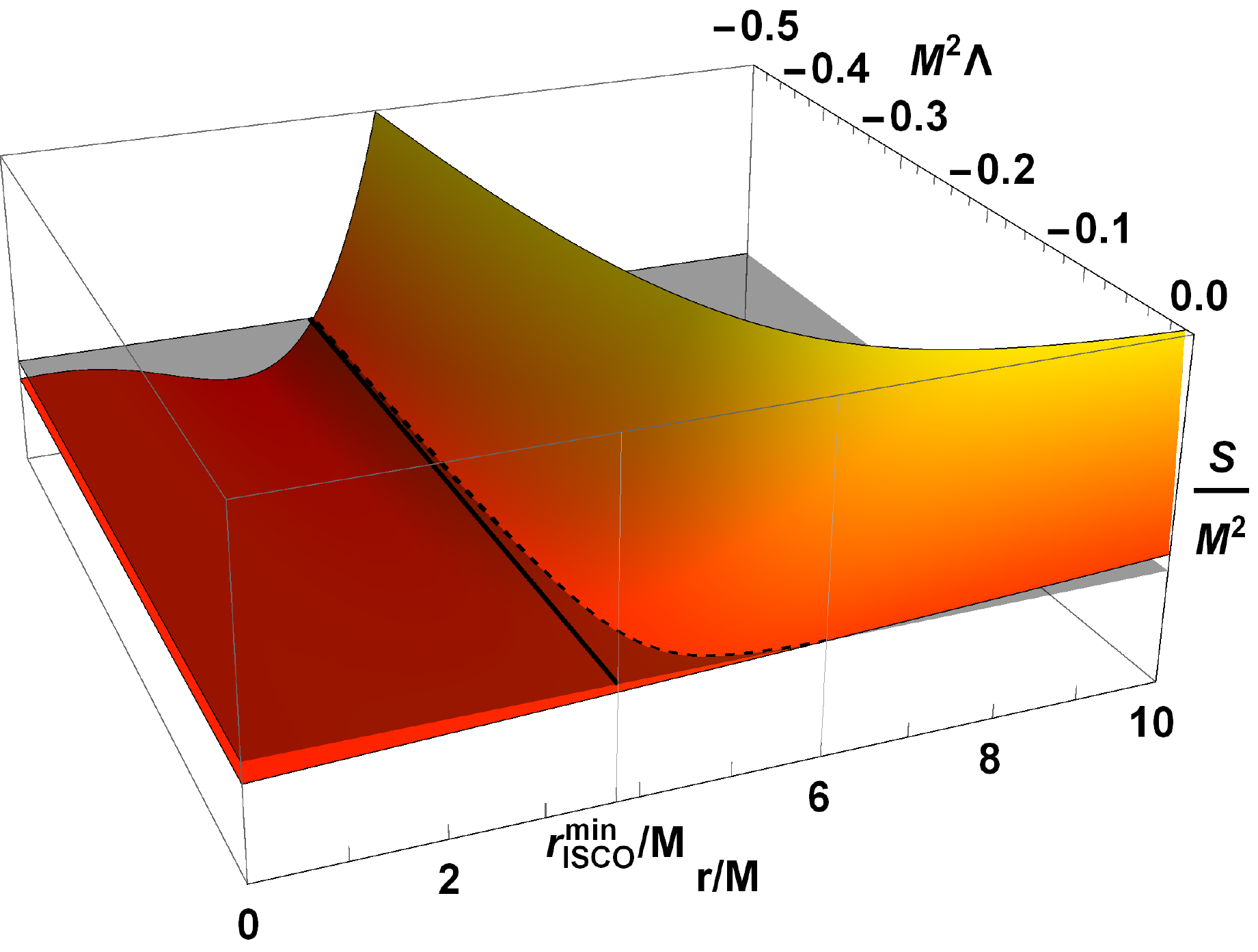}
\caption{The function $S(\Lambda,r),$ defined in Eq.~(\ref{stable_condition}). The dashed curve is the intersection of the surface $S$ with the gray plane ($S=0$). The solid line is the asymptote of the dashed curve along $\Lambda.$}
\label{fig_F_stability}
\end{figure}

In Fig.~\ref{fig_F_stability}, we show the surface function $S(\Lambda, r)$ defined in Eq.~\eqref{stable_condition}. The stable orbits are associated with the parameters $\Lambda$ and $r$ of the surface on and above the hatched plane [$S(\Lambda,r) \geq 0$].
We see that, as the value of $\Lambda$ decreases, the radial position of the \textit{innermost stable circular orbit}, $r_{ISCO}$, also decreases, since it is associated to the locus $S(\Lambda,r) = 0$ (dashed curve).
As $\Lambda \rightarrow -\infty,$ we have $r_{ISCO} \rightarrow r^{min}_{ISCO} \equiv 15 M/4$ (solid line). For $\Lambda = 0$, we have $S(\Lambda,r) \geq 0$ for $r \geq 6M$ and $r_{ISCO}=6M$.

The angular velocity of a massive particle moving along circular geodesics, as seen by the static observer at $r_{obs},$ is given by
\begin{equation}
\label{angular_velocity}
\Omega \equiv \frac{d \phi}{dt} = \frac{\dot{\phi}}{\dot{t}} = \sqrt{\frac{M}{r^3} - \frac{\Lambda}{3}},
\end{equation}
which, for $\Lambda<0,$ tends to a minimum positive constant value , $\Omega_{min} = \sqrt{-\Lambda/3}$, as the orbit radial position $r \rightarrow \infty$.
The angular velocity $\Omega$ tends to a maximum value, $\Omega_{max}$, 
as $r \rightarrow 3M$.
We obtain the power of scalar radiation emitted by the source in terms of observable quantities such as $\Omega$, $M$ and $\Lambda$. One can invert Eq.~\eqref{angular_velocity} to write the radial position $r$ of the circular orbit as a function of $\Omega$ and $\Lambda.$

In asymptotically AdS spacetimes, we must specify, beyond the initial configuration, suitable boundary conditions at infinity~\cite{hawking_1983} in order to have a well defined field theory. In the next section, we analyze the massless scalar field in the SAdS spacetime. To canonically quantize the field, we follow the prescription similar to that of Schwarzschild~\cite{birrel_1982,parker_2009,boulware_1975,
ashtekar_1975} and SdS spacetimes~\cite{higuchi_1987,brito_2020}.
\section{Scalar field in Schwarzschild--anti--de Sitter}
\label{sec_scalar_radiation}
We consider a minimally coupled massless real scalar field $\Phi(x).$ Its dynamics in a curved background can be derived from the following action:
\begin{equation}
\label{action}
S =-\frac{1}{2}\int \nabla_{\mu} \Phi(x) \nabla^{\mu} \Phi(x) \sqrt{-g} \, d^4x.
\end{equation}
The equation of motion can be written as
\begin{equation}
\label{KG}
\nabla_{\mu}\nabla^{\mu}\Phi(x) \equiv \frac{1}{\sqrt{-g}}\partial_{\mu} \left( \sqrt{-g} g^{\mu \nu} \partial_{\nu} \Phi \right) = 0,
\end{equation}
where $g = - r^4 \sin^2 \theta$ is the determinant of the SAdS metric.

\begin{figure}
\includegraphics[scale=0.42]{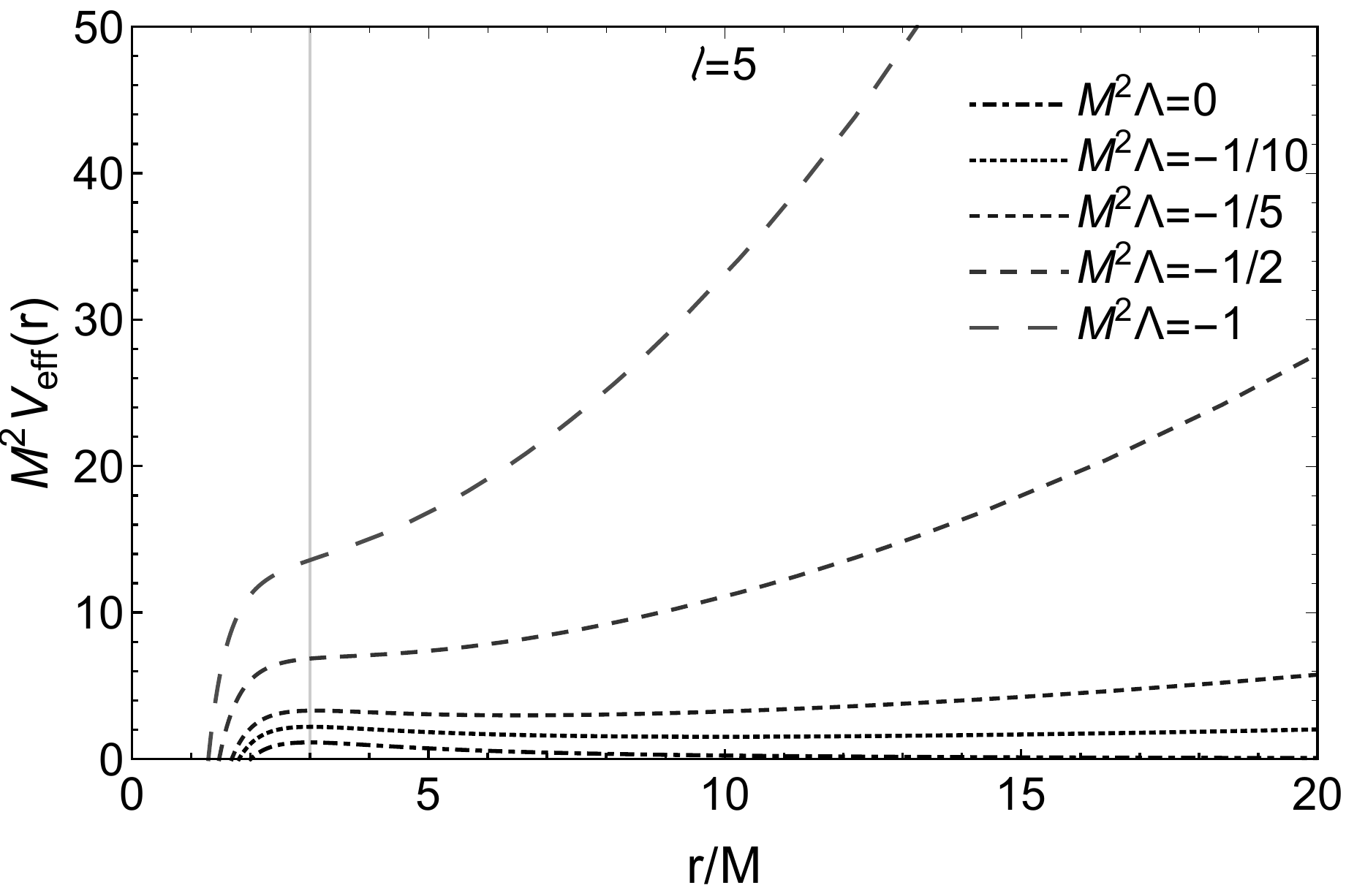}
\caption{The effective potential $M^2V_{eff},$ defined in Eq.~(\ref{effective_potential}), with $l=5$ and five different values of the cosmological constant $M^2\Lambda.$ The vertical gray line denotes the radial position of the photon sphere.}
\label{fig_effective_potential}
\end{figure}

We may define the positive-frequency modes, solutions of Eq.~\eqref{KG}, associated to the timelike Killing vector field $\partial_t,$  in the form given by
\begin{equation}
\label{Mode_positive}
u_{\omega l m}(x) = \sqrt{\frac{\omega}{\pi}} \frac{\Psi_{\omega l}(r)}{r} Y_{l m}(\theta, \phi) e^{- i \omega t} \hspace{0.4 cm} (\omega > 0),
\end{equation}
where the quantities $Y_{l m}(\theta, \phi)$ are the scalar spherical harmonics~\cite{NIST_handbook}. Using Eq.~\eqref{Mode_positive} and the eigenvalue equation of $Y_{l m}(\theta, \phi),$ it is possible to simplify Eq.~\eqref{KG} to the ordinary differential equation:
\begin{equation}
\label{dif_radial}	
\left[ - f_{\Lambda}\frac{d}{dr} \left( f_{\Lambda}\frac{d}{dr}\right) + V_{eff}\right]\Psi_{\omega l}(r) = \omega^2 \Psi_{\omega l}(r),
\end{equation}
with the effective potential $V_{eff}$ defined by
\begin{equation}
\label{effective_potential}
V_{eff}(r) \equiv f_{\Lambda}(r)\left( \frac{l(l+1)}{r^2} + \frac{2M}{r^3} - \frac{2 \Lambda}{3} \right).
\end{equation}
The potential \eqref{effective_potential} is illustrated in Fig.~\ref{fig_effective_potential}. We see that $V_{eff}(r)$ vanishes at the horizon $r_{e}$ and diverges at infinity. The potential also increases as the parameter $\Lambda$ decreases from $0$. For sufficiently small values of $\abs{\Lambda}$ (or sufficiently large values of $l$) the potential presents points of local maximum and minimum.
In contrast, the potential in the Schwarzschild spacetime presents only a point of maximum near the radial position of the circular null geodesic.
The local minimum of the effective potential in the SAdS spacetime gives rise to quasibound states, which are long-lived quasinormal modes~\cite{grain_2006,festuccia_2009,daghigh_2009,berti_2009}.

Expanding the potential $V_{eff}(r),$ defined in Eq.~\eqref{effective_potential}, near infinity, its leading order asymptotic approximation can be given by
\begin{equation}
\label{potential_at_infinity}
V_{eff}(r \rightarrow \infty) \sim \frac{2 \Lambda ^2 r^2}{9}-\frac{1}{3} \Lambda  (l (l+1)+2).
\end{equation}

In principle, we can derive asymptotic solutions of Eq.~\eqref{dif_radial} using Eq.~\eqref{potential_at_infinity}. However, in order to analyze the behavior of the field at infinity, we perform a change of variable in the radial coordinate valid outside the horizon [$r \in (r_e,+\infty) \rightarrow \rho \in (- \infty, \infty$)], given by~\cite{winstaley_2001}
\begin{equation}
\label{R_coordenate}
\rho \equiv \ln(r-r_e),
\end{equation}
together with
\begin{equation}
\label{fomega}
\Psi_{\omega l}(r) = \sqrt{\frac{r-r_e}{f_{\Lambda}(r)}} \psi_{\omega l}(\rho).
\end{equation}
From Eq.~\eqref{R_coordenate}, we see that $\rho \rightarrow -\infty$ as $r \rightarrow r_e$ and $\rho \rightarrow +\infty$ as $r \rightarrow +\infty$. Substituting Eqs.~\eqref{R_coordenate} and \eqref{fomega} into Eq.~\eqref{dif_radial}, the Klein--Gordon equation becomes
\begin{equation}
\label{dif_with_rho}
\frac{d^2}{d \rho^2}\psi_{\omega l}(\rho) + V(r)\psi_{\omega l}(\rho) = 0,
\end{equation}
with the potential function $V(r)$ defined by
\begin{equation}
\label{potential_for_dif_with_rho}
V(r) \equiv  - \frac{1}{4} + \frac{\left(r-r_e\right){}^2}{f_{\Lambda }(r){}^2} \left(\omega^2 - V_{\text{eff}}(r) - \frac{\chi}{4} \right),
\end{equation}
where
\begin{equation}
\label{function_defined_2}
\chi \equiv \left(2 f_{\Lambda }(r) f_{\Lambda }''(r)-f_{\Lambda}'(r){}^2 \right)
\end{equation}
and $r$ should be seen as a function of $\rho$, defined by the inversion of Eq.~(\ref{R_coordenate}).
At infinity ($r \rightarrow + \infty$), the function $V(r)$ tends to a constant value:
\begin{equation}
\label{potential_rho_inf}
V(r \rightarrow\infty) = - 9/4.
\end{equation}
At the event horizon $r_e$ ($r \rightarrow r_{e}$), the potential is approximated to
\begin{equation}
\label{potential_rho_r_e}
V(r \rightarrow r_e) \sim \frac{\omega^{2}}{f'_{\Lambda}(r_{e})^{2}} \equiv \bar{\omega}^2.
\end{equation}
With Eqs.~\eqref{potential_rho_inf} and \eqref{potential_rho_r_e}, we obtain the asymptotic solutions to Eq.~\eqref{dif_with_rho}. Near the event horizon, the solution is expressed by
\begin{equation}
\label{asymp_rho_re}
\psi_{\omega l} (\rho) \sim A_{\omega l} e^{i \bar{\omega} \rho} + B_{\omega l} e^{- i \bar{\omega} \rho},
\end{equation}
where the modes with coefficients $A_{\omega l}$ correspond to flux incoming from the past event horizon $H_e^-$ and those with coefficients $B_{\omega l}$ correspond to the flux reflected back down to the future event horizon $H_e^+.$
 
At infinity, we can write the solution as
\begin{equation}
\label{asymp_rho_inf}
\psi_{\omega l} (\rho) =\frac{C_{\omega l}}{\sqrt{2}} \left( e^{\frac{3}{2} \rho} - i e^{-\frac{3}{2} \rho} \right) + \frac{D_{\omega l}}{\sqrt{2}} \left( e^{\frac{3}{2} \rho} + i e^{-\frac{3}{2} \rho} \right),
\end{equation}
where the modes with coefficients $C_{\omega l}$ are associated to outgoing flux at infinity and those with coefficients $D_{\omega l}$ are associated to incoming flux (see Ref.~\cite{winstaley_2001}). We impose reflective boundary conditions at infinity, which means that the radial flux must be zero. This condition is satisfied if $|C_{\omega l}|=|D_{\omega l}|,$ i.e. if $C_{\omega l}=D_{\omega l}$ or $C_{\omega l}=-D_{\omega l}.$
Hence, there are two linearly independent solutions to Eq.~\eqref{dif_with_rho}, with asymptotic forms given by
\begin{equation}
\label{asymptotic-1}
\psi^1_{\omega l} \sim \mathcal{A}^1_{\omega l} \begin{cases}
 e^{i \bar{\omega} \rho} + \mathcal{B}^1_{\omega l}e^{-i\bar{\omega} \rho} , & \rho \rightarrow -\infty, \\
 \mathcal{N}^1_{\omega l} e^{\frac{3}{2} \rho}, & \rho \rightarrow +\infty,
\end{cases}
\end{equation}
and
\begin{equation}
	\label{asymptotic-2}
	\psi^2_{\omega l} \sim \mathcal{A}^2_{\omega l} \begin{cases}
		e^{i \bar{\omega} \rho} + \mathcal{B}^2_{\omega l}e^{-i\bar{\omega} \rho} , & \rho \rightarrow -\infty, \\
		\mathcal{N}^2_{\omega l} e^{-\frac{3}{2} \rho}, & \rho \rightarrow +\infty,
	\end{cases}
\end{equation}
where $\mathcal{A}^k_{\omega l} $ (with $k=1,2$) are overall normalization constants.
The constants $\mathcal{N}^k_{\omega l}$ and $\mathcal{B}^k_{\omega l}$ are complex quantities. Note that we have defined the basis modes such that they are associated with a unitary flux coming up out of $H_e^-$.
By considering the Wronskian of the solutions, given by Eqs.~\eqref{asymptotic-1} and \eqref{asymptotic-2}, we can show that $\abs{\mathcal{B}^{k}_{\omega l}} = 1,$ due to conservation of the flux.
The function $\psi^{1}_{\omega l}$ diverges as $\rho \to \infty$ or, equivalently, $r \to \infty$. However, the actual scalar mode $\Psi^{1}_{\omega l m}/r$ is proportional to $R_{{AdS}}$ as $r \to \infty$.

To normalize the modes $u_{\omega l m},$ we use the inner product defined by
\begin{equation}
\left(\Psi_1, \Psi_2 \right) \equiv i \int_{\Sigma} d\Sigma^{\mu}\left( \Psi_1^* \left(\nabla_{\mu} \Psi_2 \right) - \left( \nabla_{\mu} \Psi_1^* \right)\Psi_2 \right),
\label{inner_product}
\end{equation}
where $d\Sigma^{\mu} = d\Sigma n^{\mu},$ with $n^{\mu}$ being a future directed unit vector orthogonal to a hypersurface $\Sigma$, defined by $t=\text{constant}$~\cite{hawking_1973, parker_2009}. It can be shown that this inner product does not depend on the choice of hypersurface if $\Psi_{1}$ and $\Psi_{2}$ satisfy the field equations~\cite{parker_2009}.
In globally hyperbolic spacetimes, such as the Kerr--Newman spacetime family, the surface $\Sigma$ is a Cauchy surface.

Imposing the orthogonality conditions
\begin{equation}
\label{u_orthogonality}
\left(u_{\omega l m},u_{\omega' l' m'} \right) = \delta_{l l'} \delta_{m m'} \delta(\omega - \omega')
\end{equation}
and
\begin{equation}
\label{u_ortogonality_null}
\left( u_{\omega l m},u^{*}_{\omega' l' m'} \right) = \left( u^{*}_{\omega l m},u_{\omega' l' m'} \right) = 0,
\end{equation}
we can obtain the absolute value of the overall normalization constants $\mathcal{A}^k_{\omega l},$ namely
\begin{equation}
\label{normali_A_1}
\abs{\mathcal{A}^k_{\omega l}} = \frac{1}{2 \sqrt{\omega \bar{\omega}}}.
\end{equation}

We canonically quantize the scalar field following the prescription for a static spacetime~\cite{birrel_1982,parker_2009,higuchi_1987,
ashtekar_1975}, in which one expands the quantum field operator $\hat{\Phi}(x)$ in terms of negative and positive frequencies associated to creation  ($\hat{a}^{\dagger}_{\omega l m}$) and annihilation ($\hat{a}_{\omega l m}$) operators. The quantum field is thus given by
\begin{equation}
\label{field_expansion}
\hat{\Phi}(x) = \sum_{l,m, k} \int_0^{\infty}  \left[u^k_{\omega l m}(x)\hat{a}^k_{\omega l m} + u^{k *}_{\omega l m}(x) \hat{a}^{k \dagger}_{\omega l m} \right] d\omega,
\end{equation}
where the index $k=1,2$ stands for modes associated to the two solutions given by Eqs.~\eqref{asymptotic-1} and \eqref{asymptotic-2}. The creation and annihilation operators satisfy the usual nonvanishing commutation relations, namely
\begin{equation}
\label{comutation_relation}
\left[\hat{a}^k_{\omega l m},\hat{a}^{k \dagger}_{\omega' l' m'} \right] = \delta_{l l'} \delta_{m m'} \delta(\omega - \omega').
\end{equation}
The vacuum state $|0\rangle$ can be defined as the quantum state that vanishes when acted upon by all annihilation operators~\cite{fulling_1973}. Thus,
\begin{equation}
\label{vacuum}
\hat{a}^k_{\omega l m} \ket{0} \equiv 0, \hspace{1 cm} \forall \hspace{0.3 cm} (\omega, l, m, k),
\end{equation}
and the one-particle-states are obtained by applying a creation operator on the vacuum state, that is
\begin{equation}
\label{one_parti_state}
\ket{k;\omega l m} = \hat{a}^{k \dagger}_{\omega l m} \ket{0}.
\end{equation}

In the next section, we compute the one-particle-emission amplitude by considering the massless scalar field coupled to a classical matter source in a timelike geodesic circular orbit on the equatorial plane of a SAdS black hole.

\section{Scalar radiation and emitted power}
\label{Sec_radiation}
The orbiting scalar source at $r=R,$ with angular velocity $\Omega$, is described by the normalized current
\begin{equation}
\label{current}
j(x) = \frac{\varrho}{\sqrt{-g} u^0} \delta(r-R) \delta(\theta - \pi/2)\delta(\phi - \Omega t),
\end{equation}
such that $\int d\beta^{(3)}j(x) = \varrho,$ where $\beta^{(3)}$ is a hypersurface orthogonal to the source's $4-$velocity $u^{\mu},$ and $\varrho$ is the magnitude of the scalar source.
The associated $4-$velocity is given by
\begin{equation}
\label{four_velocity}
u^{\mu}(R) = u^0 \left(1,0,0,\Omega \right),
\end{equation}
with $u^0$ fixed by the condition $u^{\mu}u_{\mu} = -1.$ Therefore,
\begin{equation}
\label{gamma_factor}
u^0 = \left( f_{\Lambda}(R) - R^2 \Omega^2 \right)^{-1/2} = \left( 1 - \frac{r_0}{R} \right)^{-1/2}.
\end{equation}

The interaction between the field and the source is carried out by adding to the action \eqref{action} the following term:
\begin{equation}
\label{interaction_action}
\hat{S}_{I} = \int d^4x \sqrt{-g} j(x) \hat{\Phi}(x),
\end{equation} 
so that the scalar source $\varrho$ can be regarded as a coupling constant that determines the magnitude of the interaction.

There is a nonvanishing probability for the source to radiate scalar quanta. To first order in perturbation theory, the transition amplitude from the vacuum-state, defined in Eq.~(\ref{vacuum}), to the one-particle-state, with energy $\omega$ and quantum numbers $l,$ $m,$ expressed in Eq.~\eqref{one_parti_state}, is written as~\cite{itzykson_1980}
\begin{equation}
\label{probability_amplitude}
A^{k; \omega l m}_{em} = \bra{k; \omega l m} i \hat{S}_I \ket{0} = i \int d^4x \sqrt{-g} j(x)u^{k *}_{\omega l m}(x).
\end{equation}
As can be seen by evaluating the integral, the probability amplitude, given by Eq.~(\ref{probability_amplitude}), is proportional to $\delta(\omega - m\Omega),$ i.e. the emitted scalar particles have frequency proportional to the angular velocity of the source ($\omega_m \equiv m\Omega$).\footnote{Although we are considering only lowest order in perturbation theory, non-perturbative calculations in Minkowski spacetime for the scalar radiation emitted by a uniformly accelerated source show that only zero-energy Rindler particles compose the final state~\cite{Landulfo2019,Fulling2020}.
The present case shares most of its technical details with the uniformly accelerated source case presented in~\cite{Landulfo2019}. Thus, we might expect that only particles obeying the condition $\omega = m \Omega$ will be emitted.}
Furthermore, there is no emission of observable scalar particles with $m=0.$ We note that $\omega$ and $\Omega$ are positive quantities, hence we must have $m \geq 1.$
The emitted power, with fixed $l$ and $m$ for each mode $k$, is given by
\begin{equation}
\label{partial_power_implicit}
W^{k;l m}_{em} = \int_{0}^{\infty} d\omega \ \omega \frac{\abs{A^{k; \omega l m}_{em}}^2}{T},
\end{equation}
where $T=\int dt = 2 \pi \delta(0)$ is the total time during which the source radiates, assumed to be the whole range of coordinate time  $t,$ with $-\infty < t < \infty$~\cite{breuer_1975,crispino_1998}.

With Eqs.~\eqref{probability_amplitude} and \eqref{partial_power_implicit}, we readily obtain the partial emitted power, namely
\begin{equation}
\label{partial_power}
W^{l m}_{em} = 2 \varrho^2 \omega_m^2 \gamma \left( \abs{\psi^1_{\omega_m l}}^2 + \abs{\psi^2_{\omega_m l}}^2\right) \abs{Y_{l m}\left(\frac{\pi}{2}, 0 \right)}^2,
\end{equation}
with
\begin{equation}
\label{GR_factor}
\gamma \equiv \left( 1 - \frac{r_0}{R} \right)\frac{R-r_e}{R^2 f_{\Lambda}(R)}.
\end{equation}
From Eq.~\eqref{GR_factor}, we see that $W^{l m}_{em} \rightarrow 0$ as $R \rightarrow r_0$ and $R \rightarrow \infty.$
Furthermore, there is no emission for odd values of $l+m,$ because the time independent quantity $\abs{Y_{l m}(\pi/2, 0)}$ vanishes in this case. For even values of $l+m,$ we can write~\cite{gradshteyn_1980}
\begin{equation}
\label{harmonic_even_values}
\abs{Y_{l m}\left(\frac{\pi}{2}, 0 \right)}^2 = \frac{2l+1}{4\pi}\frac{(l+ m - 1)!!(l - m - 1)!!}{(l+m)!!(l-m)!!}.
\end{equation}

The total emitted power $W_{em}$ is obtained by summing the contributions of all multipoles, namely
\begin{equation}
\label{total_power}
W_{em} = \sum_{l=1}^{\infty} \sum_{m=1}^{l} W^{l m}_{em}.
\end{equation}

In order to obtain the emitted power, we solve numerically the differential equation for $\psi^k_{\omega l},$ given by ~\eqref{dif_with_rho}. We follow a procedure similar to that stated in Schwarzschild and SdS spacetimes. In the next section, we present some representative results.

\begin{figure*}
\centering
\includegraphics[scale=0.45]{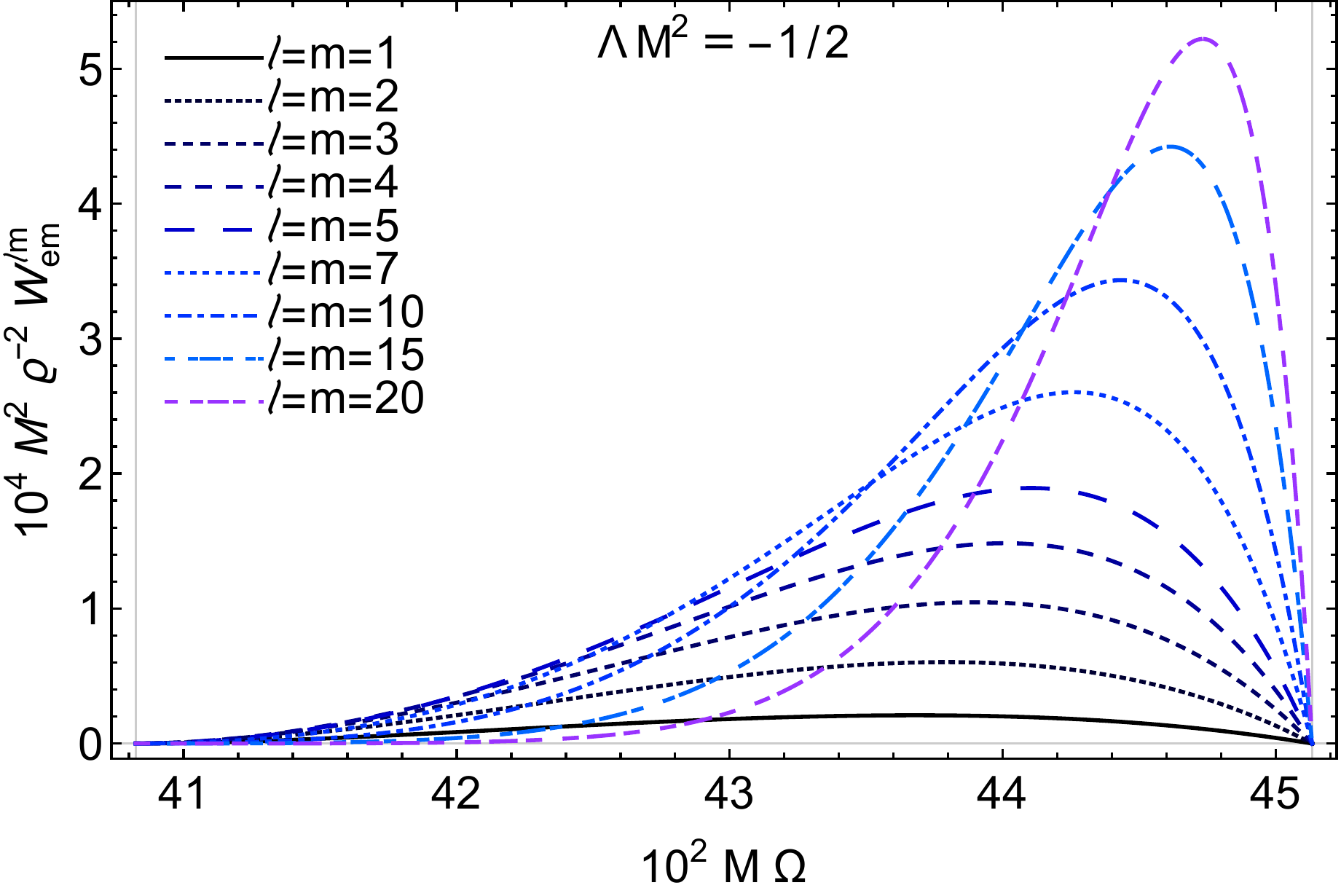} \hspace{0.3 cm}
\includegraphics[scale=0.45]{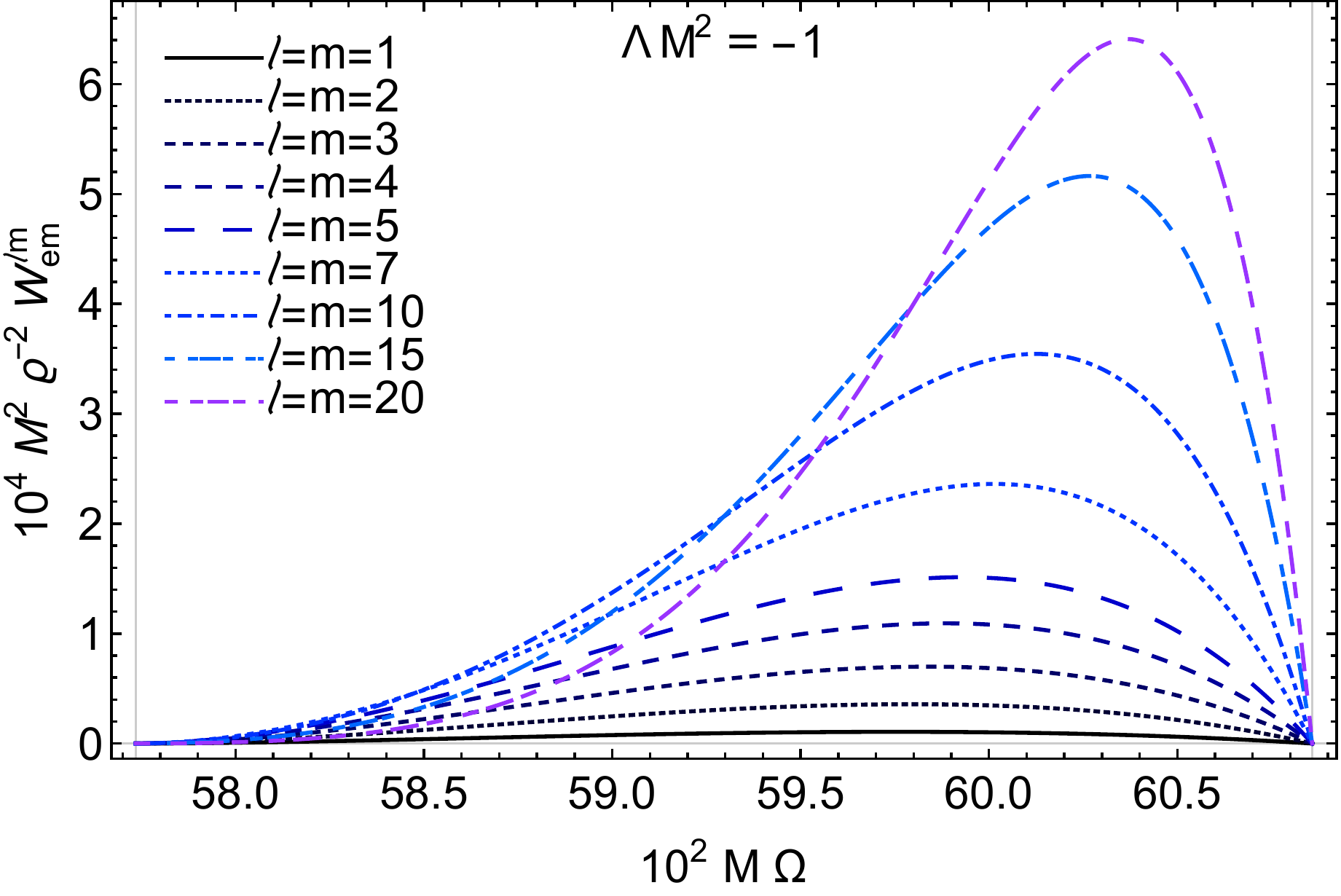}
\includegraphics[scale=0.45]{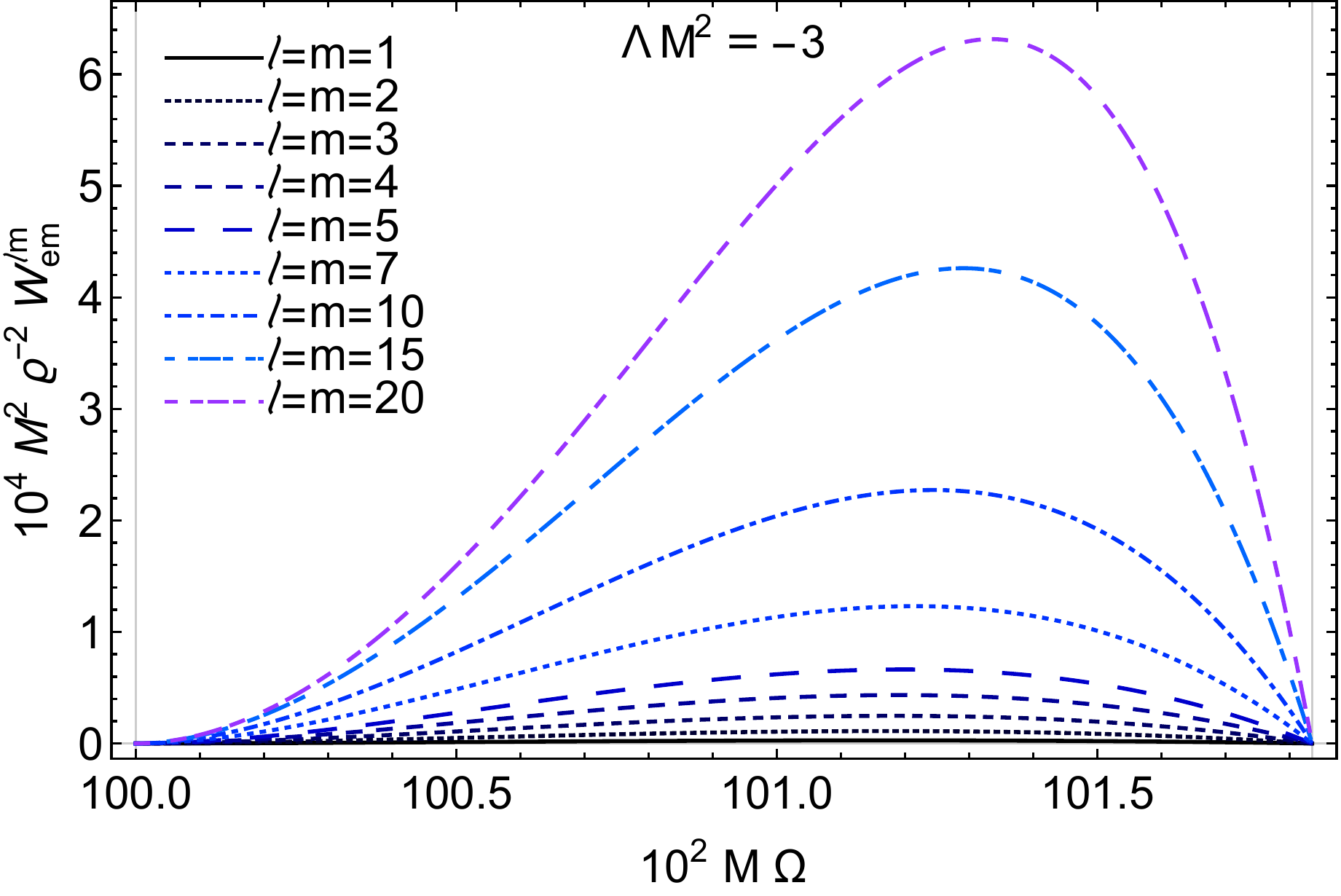}
\hspace{0.3 cm}
\includegraphics[scale=0.45]{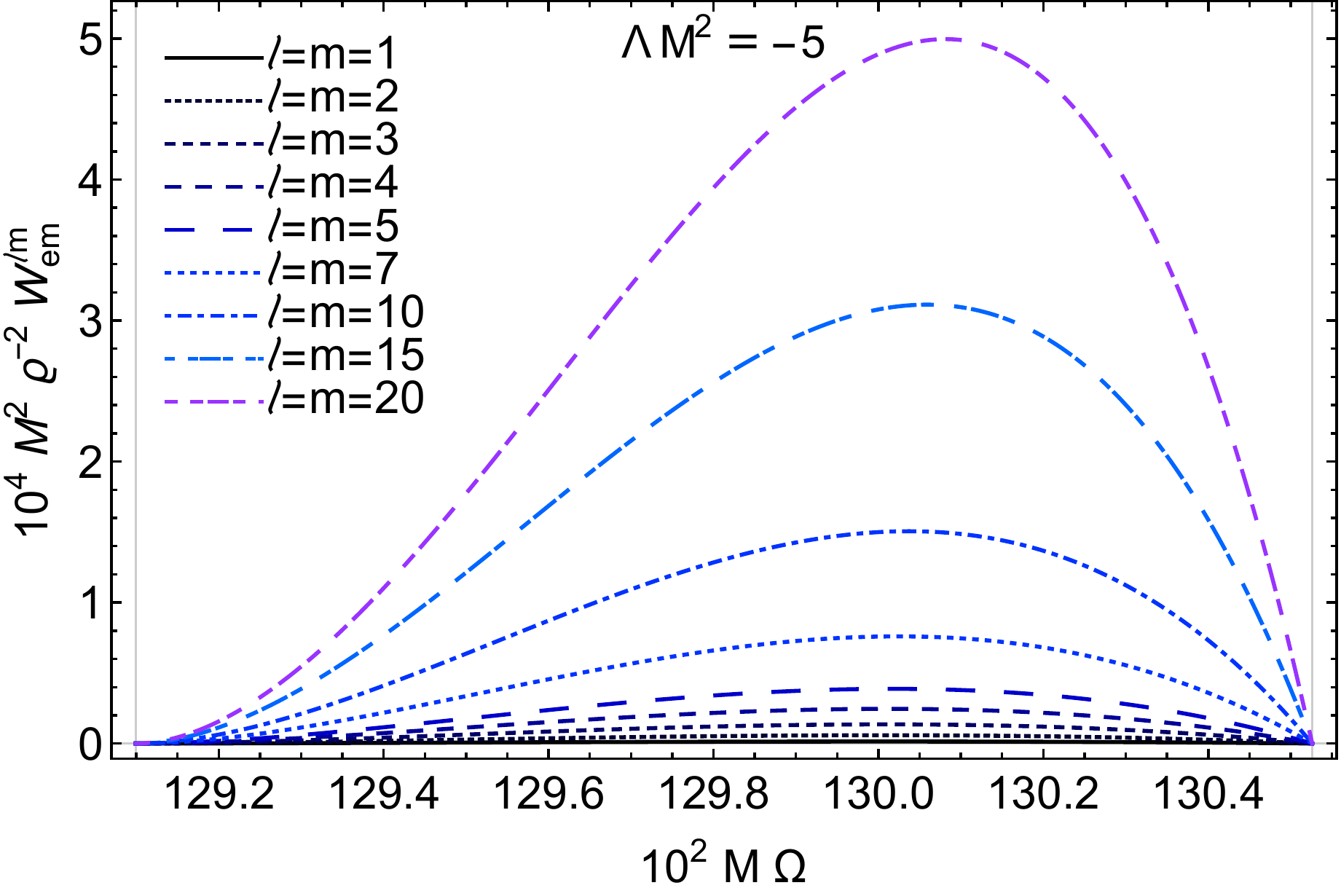}
\caption{The emitted power as a function of $M\Omega,$ given by Eq.~\eqref{partial_power}, for $M^2 \Lambda = -2^{-1}$ (top-left), $M^2 \Lambda = -1$ (top-right), $M^2 \Lambda = -3$ (bottom-left), $M^2 \Lambda = -5$ (bottom-right) and different choices of $l=m,$ as indicated. The curves are plotted from $\Omega_{min},$ corresponding to the radial position $R \rightarrow + \infty,$ up to the value $\Omega_{max},$ corresponding to the radial position $R=3M$.}
\label{Pot_Parc_SAdS_Lamb1}
\end{figure*}

\section{Results for the emitted power}
\label{sec_results}

Using methods outlined in Refs.~\cite{bernar_2017,brito_2020}, we compute numerically the scalar power emitted by the source for different choices of the parameters. We start with the integration of Eq.~\eqref{dif_with_rho}, using boundary conditions given by Eqs.~\eqref{asymptotic-1} and \eqref{asymptotic-2}. The integration is performed in the region $r_e + \delta^{-1}r_e \leq r \leq \delta r_e,$ where $\delta \equiv 10^{5}.$ We compute the coefficients $\abs{\mathcal{N}^{k}_{\omega l}}$ by matching the numerically obtained solutions with the asymptotic forms, given by Eqs.~\eqref{asymptotic-1} and \eqref{asymptotic-2}, near the event horizon.

Note that we are free to choose either $M$ or $R_{AdS}$ equal to the unity. This choice corresponds to rescaling the radial coordinate as $r \rightarrow r/M$ or $r\rightarrow r/R_{AdS},$ respectively. In this section, we display our results considering $M$ as the fundamental scale, making use of the $R_{AdS}$ scale when convenient.
We shall analyze the radiation scenario in which $M^2\Lambda \leq -1/2,$ such that $r_e > 0.6 R_{AdS},$ which corresponds to a wide variety of black hole sizes, when compared to the anti--de Sitter radius.

We note that, for the values of $\Lambda$ and $l$ considered in this section, more than $98\%$ of the emitted power comes from the $l=m$ mode. This percentage decreases with decreasing $\Lambda.$

In Fig.~\ref{Pot_Parc_SAdS_Lamb1}, we show the partial emitted power for four choices of the cosmological constant $\Lambda$ and different choices of the multipole number $l=m.$ We see, as a characteristic behavior, that the power for each value of $l=m$  increases from zero at $\Omega_{min}$ ($R \rightarrow \infty$), reaches a maximum and decreases to zero as $\Omega \rightarrow \Omega_{max}$ ($R \rightarrow r_0$).

For all values of $\Lambda$ considered, we have found that the peak of the partial emitted power increases with $l=m$.
Additionally, the value of $\Omega$ where this peak is located gets closer to $\Omega_{max}$ as we increase $l=m$.
This shows the dominance of higher multipole modes in the power for orbits closer to the photon sphere.

We also see that emission of higher multipole modes is enhanced with decreasing $M^2\Lambda$. For spacetimes with $|\Lambda| \geq 3/M^{2}$, even orbits far from the black hole, i.e. stable orbits, present this behavior of increasing partial emitted power with $l=m$. We note that $|\Lambda| = 3/M^{2}$ is the point in the parameter space where $r_{e}=R_{AdS}$. Thus, the region $|\Lambda| \geq 3/M^{2}$ in the parameter space denotes the large black hole regime.
This is in agreement with the classical result reported in Ref.~\cite{cardoso_2002}, where the Green's function approach was used.
In contrast, the analogous case in SdS spacetime ($\Lambda > 0$) presents an enhancement of lower multipole modes as $M^2\Lambda$ increases~\cite{brito_2020}.

\begin{figure}[h!]
\includegraphics[scale=0.45]{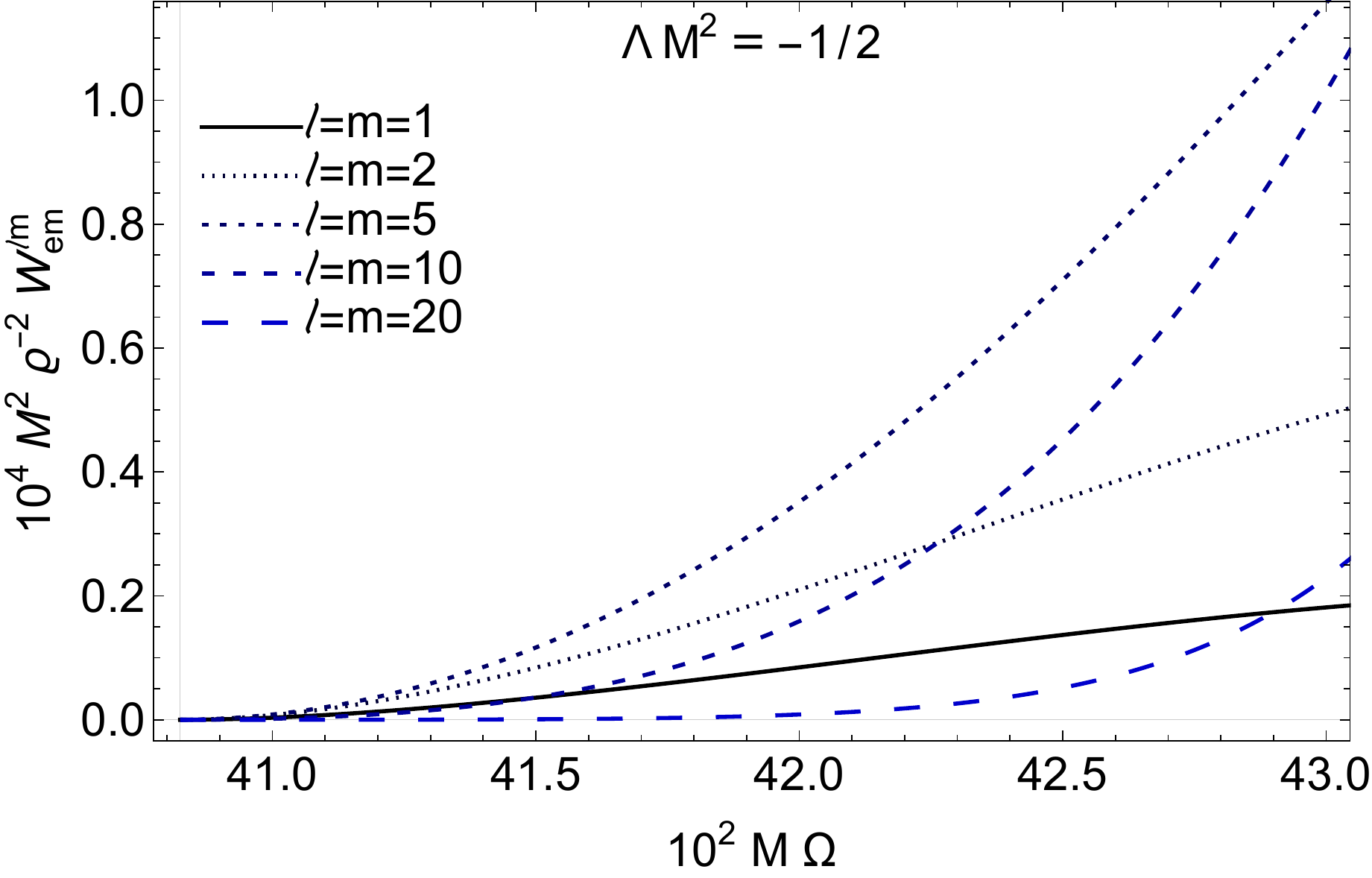}
\caption{The partial emitted power as a function of $M\Omega,$ for $M^2\Lambda = -2^{-1}$ and different choices of $l=m$. The angular velocity range is from $\Omega_{min}$ up to the maximum allowed value for stable orbits.}
\label{fig:partial_power_stable_orbits_Lamb_1_2}
\end{figure}

For $\Lambda = -1/(2M^{2}),$ the last stable circular orbit occurs at $r \approx 3.8 M$ (or $M \Omega \approx 0.43$).
We see, on Fig.~\ref{fig:partial_power_stable_orbits_Lamb_1_2}, that at this position the power is dominated by the contribution from modes with $l \geq 2$.
Furthermore, this does not occur in the Schwarzschild (cf. Fig.~2 of Ref.~\cite{crispino_2000}) or SdS (cf. Fig.~7 of Ref.~\cite{brito_2020}) spacetime scenarios.
Numerically we find that, for a circular orbit in $r_{ISCO}$, we have $W_{em}^{2 2} \geq W_{em}^{1 1}$ when $M^2 \Lambda \lesssim - 0.0791$.

\begin{figure}[h!]
\includegraphics[scale=0.45]{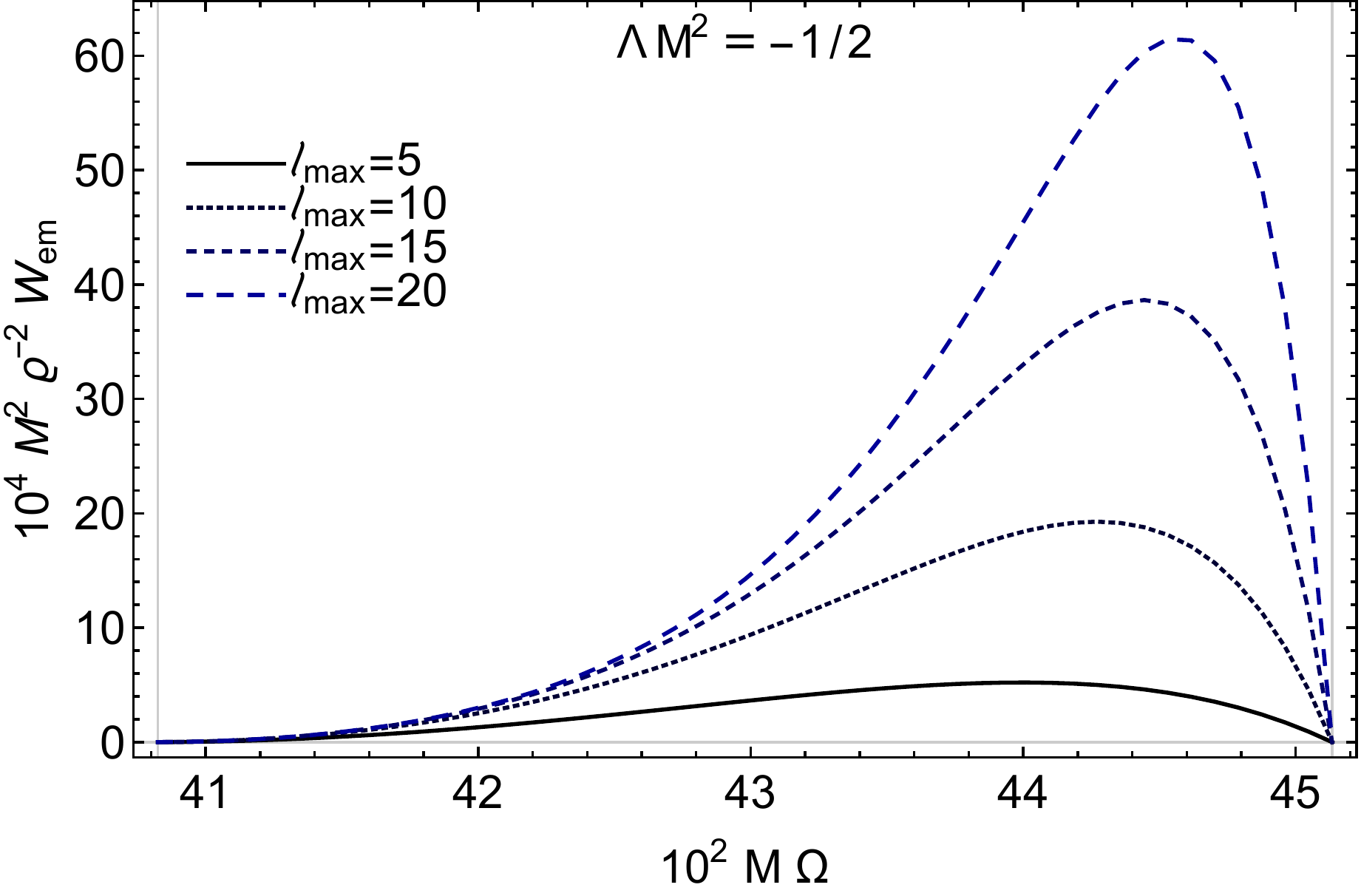}
\caption{The total emitted power as a function of $M\Omega,$ given by Eq.~\eqref{total_power}, for $M^2\Lambda = -2^{-1}$ and different choices of $l_{max},$ as indicated.}
\label{Fig_Total_Power}
\end{figure}

In Fig.~\ref{Fig_Total_Power}, we show the total emitted power, where the characteristic synchrotronic behavior can be readily observed for unstable orbits ($M \Omega > 0.43$).
The summation in $l$ was truncated at a given $l=l_{max}$.
We can see that, as we include the contribution from higher multipoles to the emitted power (by changing $l_{max}$), the total emitted power increases. For unstable orbits, a substantial portion of the multipoles higher than the fundamental dipole mode have non negligible contributions to the emitted power. We note, however, that for any finite value $R < 3M$ of the source's radial position, the multipole contributions to the total emitted power are negligible past a certain $R$-dependent value of $l$. This also happens in asymptotically flat spacetimes.
In fact, our numerical results suggest that sufficiently high multipole modes do not contribute significantly to the total emitted power, as can be seen in Figs.~\ref{Pot_Parc_SAdS_Lamb1} and~\ref{fig:partial_power_stable_orbits_Lamb_1_2}, namely: There is an initial enhancement in the contribution of increasing multipole numbers $l,$ but there is a maximum $l$ beyond which the contribution to the emitted power starts to decrease. The peak in the total emitted power, plotted in Fig.~\ref{Fig_Total_Power}, approaches $R = 3M$ as we sum higher multipole contributions, analogously to what happens in asymptotically flat spacetimes. This occurs because closer to the black hole photon sphere, the scalar source emits radiation of the synchrotron type, i.e, the contribution of higher multipole modes becomes dominant as $R \rightarrow 3M$ ($\Omega \rightarrow \Omega_{max}$).
In Fig.~\ref{Fig_Total_Power}, the peaks in the total power occur at $10^2 M \Omega \approx \{43.99,44.27,44.44,44.56\}$ for $l_{max} = \{5, 10, 15, 20 \},$ respectively.

\section{Final Remarks}
\label{Sec_remarks}
We have investigated the synchrotron scalar radiation in the Schwarzschild--Anti--de Sitter (SAdS) spacetime using the semiclassical approach of quantum field theory in curved spacetimes at tree level.
We have presented results for the power of scalar radiation emitted by the source as a function of its angular velocity $\Omega$ for some values of the cosmological constant $\Lambda$. 
In the region of the parameter space analyzed in the present investigation, we have obtained good agreement between our semiclassical results and the classical ones previously reported in the literature~\cite{cardoso_2002}.

Different values of $\abs{\Lambda}$ can be associated to different black hole sizes, when compared to the anti-de Sitter radius. Given that the SAdS spacetime is not globally hyperbolic, we must impose suitable boundary conditions at spatial infinity. We considered reflective boundary conditions such that there is no net radial flux at the AdS boundary. This gives rise to two linearly independent sets of modes, which are normalizable and can be used to construct a quantum theory for the massless scalar field. The emitted power can be obtained by computing the one-particle emission amplitude when the quantum scalar field is excited by the source in a geodesic circular orbit.


Our results show that for $M^2 \Lambda \leq -1/2,$ there is an enhancement in the contribution of the high multipole modes to the emitted power. The radiation exhibits a synchrotron-like behavior for orbits much farther from the black hole, when compared to the corresponding case in Schwarzschild spacetime. For sufficiently large $\abs{\Lambda}$, this behavior happens even for stable circular geodesics.
Results in the Schwarzschild--de Sitter (SdS) spacetime case show the opposite behavior. There is an enhancement in the contribution of the lower multipole modes as $M^2 \Lambda$ increases~\cite{brito_2020}.

\begin{acknowledgments}
The authors thank Funda\c{c}\~ao Amaz\^onia de Amparo a Estudos e Pesquisas (FAPESPA),  Conselho Nacional de Desenvolvimento Cient\'ifico e Tecnol\'ogico (CNPq) and Coordena\c{c}\~ao de Aperfei\c{c}oamento de Pessoal de N\'{\i}vel Superior (Capes) - Finance Code 001, in Brazil, for partial financial support.
This work has been further supported by the European Union's Horizon 2020 research and innovation (RISE) programme H2020-MSCA-RISE-2017 Grant No.~FunFiCO-777740.
\end{acknowledgments}


\end{document}